\documentclass[preprint,prb,showpacs,amsmath,amssymb]{revtex4}
\usepackage{graphicx}
\usepackage{dcolumn}
\usepackage{bm}
\usepackage{amsfonts}
\usepackage{amsmath}
\usepackage{amssymb}

\begin{document}
\title{Hysteretic ac losses in a superconductor strip between flat magnetic shields}
\author{Y.A.~Genenko}
\email{yugenen@tgm.tu-darmstadt.de}
\author{H.~Rauh} 

\affiliation{Institut f\"ur Materialwissenschaft, Technische Universit\"at Darmstadt, 
64287 Darmstadt, Germany}%
\date{\today}

\begin{abstract}
Hysteretic ac losses in a thin, current-carrying superconductor strip located between
two flat magnetic shields of infinite permeability are calculated using Bean's model 
of the critical state. For the shields oriented parallel to the plane of the strip, 
penetration of the self-induced magnetic field is enhanced, and the current dependence 
of the ac loss resembles that in an isolated superconductor slab, whereas for the 
shields oriented perpendicular to the plane of the strip, penetration of the 
self-induced magnetic field is impaired, and the current dependence of the ac loss is 
similar to that in a superconductor strip flanked by two parallel superconducting 
shields. Thus, hysteretic ac losses can strongly augment or, respectively, wane when 
the shields approach the strip.
\end{abstract}


\maketitle

\section{\label{sec:intro}Introduction}

The problem of reducing hysteretic ac losses has lately become a major issue for 
large-scale superconductor applications~\cite{GomorySUST2006,Malozemoff2009,Clem2009}. 
In planar superconductors such as thin superconductor strips, a promising way to curtail 
the dissipation of electromagnetic energy exploits the shielding effect of magnetically 
susceptible environments, thereby controlling the distributions of the transport current 
and the magnetic field~\cite{Majoros2000_1,Genenko2002_1,Genenko2004_1,AladSUST2009,%
Genenko2009}. The quest to minimize hysteretic ac losses in superconductor/magnet 
composites, on the other hand, arises naturally because of the wide practical use of 
soft-magnetic substrates for the fabrication of superconductor-coated tapes. Despite 
great effort directed at numerical simulations of composites of the above 
sort~\cite{Alamgir2005,GomoryAPL2006,Gu2007,GomorySUST2009}, a theoretical analysis 
which would allow simple estimates of the penetration of magnetic flux and the 
consequential hysteretic ac losses in thin superconductor strips for basic 
configurations like, {\it e.g.}, bilayered superconductor/magnet 
heterostructures~\cite{Mawatari2008}, has not come forth yet. Here, therefore, we 
present such calculations in the case of a magnetically shielded superconductor strip 
for two fundamental shielding geometries, {\it viz.} bulk flat magnets oriented parallel or, 
respectively, perpendicular to the plane of the strip.

\section{\label{sec:generalmodel}Theoretical model}

\begin{figure*}[tbp]
\begin{center}
    \includegraphics[width=7.5cm]{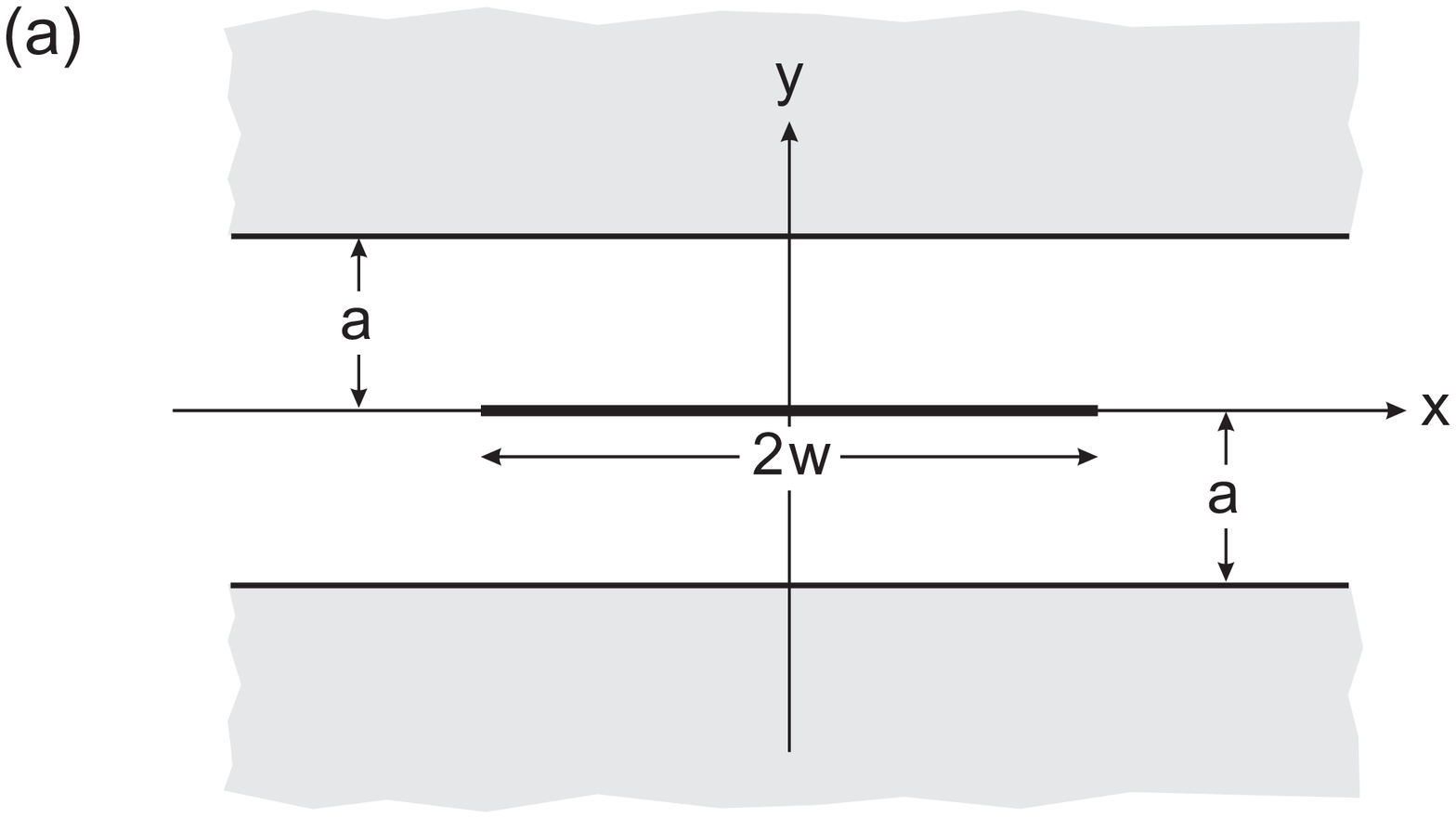}
    \hspace{1cm}
     \includegraphics[width=7.5cm]{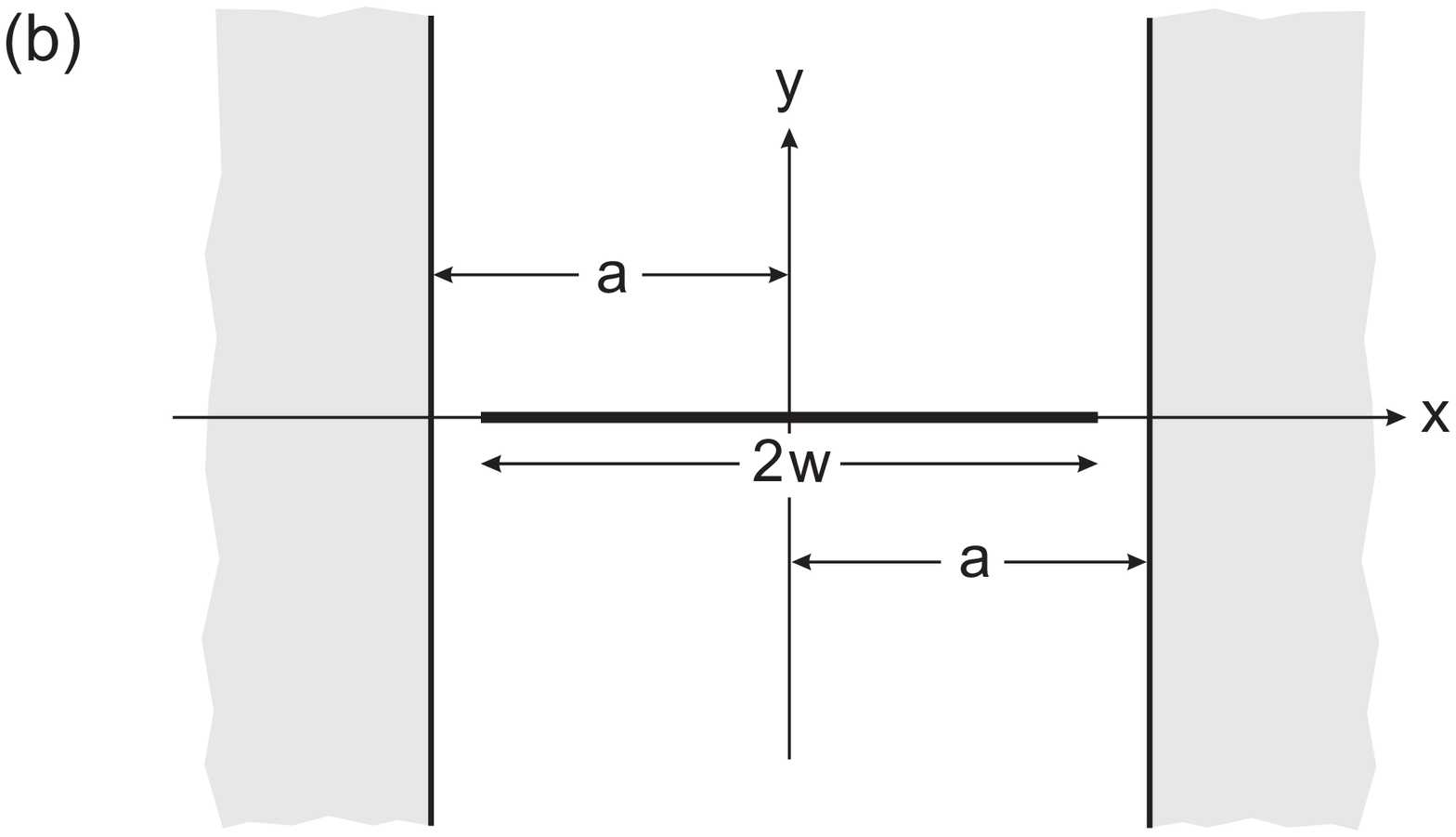}
    \caption{Cross-sectional view of a superconductor strip of width $2w$ (dark shading) 
             located between two bulk flat magnets (light shading) extending infinitely 
             in the $z$-direction of a Cartesian coordinate system $x, y, z$, adapted to 
             the strip for (a) the longitudinal shielding geometry and (b) the transverse 
             shielding geometry. The definition of the distance $a$ between the surfaces 
             of the magnets and the centre of the strip is indicated for either magnet 
             configuration.}   
\label{skizze}
\end{center}
\end{figure*}

We consider an infinitely extended type-II superconductor strip of width $2w$ and 
thickness $d\ll w$ limited by the range $-w\leq x\leq w$ and located between two 
infinitely extended, homogeneous soft magnets, the direction of translational invariance 
of this heterostructure being parallel to the $z$-axis of a Cartesian coordinate system 
$x, y, z$, with vertical or, respectively, horizontal distance $a$ between the surfaces 
of the magnets and the centre of the strip, as depicted in Fig.~\ref{skizze}. 
The magnets are understood to reveal permeability $\mu\rightarrow\infty$; an idealization
which has proven representative for real magnets with relative permeability exceeding 
about two hundred~\cite{Genenko2000}. The strip is supposed to carry a longitudinal 
transport current that changes periodically with time, at fixed amplitude $I$, in the 
absence of an externally applied magnetic field. By virtue of the restriction concerning 
the dimensions of the strip, spatial variations of the current and the self-induced 
magnetic field on a length scale less than $d$ may be ignored and, for mathematical 
convenience, the strip regarded as infinitesimally thin, enabling the physical state of 
the strip to be characterized by the sheet current $J$ as a function of $x$ alone.

Implementing Bean's model of the critical state duly adapted to the geometry of the 
strip~\cite{Norris1970,Brandt1993}, we assume that the dynamics of the magnetic flux is 
controlled by the field-independent critical sheet current $J_c=dj_c$  with the critical 
current density $j_c$. As long as the amplitude of the ac transport current $I$  stays 
below the maximum loss-free current $I_c=2wJ_c$, a flux-free region of half-width $b<w$ 
prevails in the central part of the strip,  $-b\leq x\leq b$, where the normal component of the 
magnetic field $H_n$  disappears, while in the marginal, flux-penetrated parts of the 
strip, $-w\leq x\leq b$  and  $b\leq x\leq w$, the sheet current $J$  equals $J_c$. Like 
the tangential component of the magnetic field, the sheet current is continuous over the 
width of the strip~\cite{Genenko2000}.

The scenario of entry and exit of magnetic flux in the presence of magnetic shields is 
essentially the same as for an unshielded strip~\cite{Norris1970,Brandt1993}; in 
particular, the distribution of the magnetic field along the strip, emerging during the 
gradual increase of the transport current from the virgin state of the strip up to the 
state associated with the maximum value of the current, changes sign as the current 
alternates between $I$ and $-I$. Accordingly, by resorting to a quasistatic approach, 
the energy dissipated during a cycle of the ac transport current, per unit length of 
the strip, amounts to
\begin{equation}
\label{ac-loss}
U_{ac} = 8 \mu_0 J_c \int_b^w dx \int_b^x dx' H_n(x').
\end{equation}
with the vacuum permeability $\mu_0 $.

\subsection{\label{subsec:lonq} Longitudinal shielding geometry}

Following previous analysis, for the bulk flat magnets oriented parallel to the plane of 
the strip as shown in Fig.~\ref{skizze}(a), the normal component of the magnetic field in 
the flux-filled margins of the strip, $-w\leq x\leq b$  and  $b\leq x\leq w$, 
reads~\cite{Genenko2000} 
\begin{equation}
\label{field-par}
H_n(x) =H_c\, \mbox{sgn}(x)\, \mbox{arctanh}\, \sqrt{ \frac{u^2(x)-p^2}{q^2-p^2} },   
\end{equation}
where $H_c=J_c/\pi$, apart from $u(x)=\tanh{(\pi x/2a)}$, $p=\tanh{(\pi b/2a)}$ and 
$q=\tanh{(\pi w/2a)}$. Herein, since the half-width of the flux-free zone $b$ depends on 
the transport current $I$  itself, 
\begin{equation}
\label{p(I)}
p = \sqrt{1 - \cosh^2{\left( \frac{\pi w}{2a}\frac{I}{I_c} \right)}  
\mbox{sech}^2{\left( \frac{\pi w}{2a} \right)} }.
\end{equation}

\begin{figure*}[tb]
    \includegraphics[width=8cm]{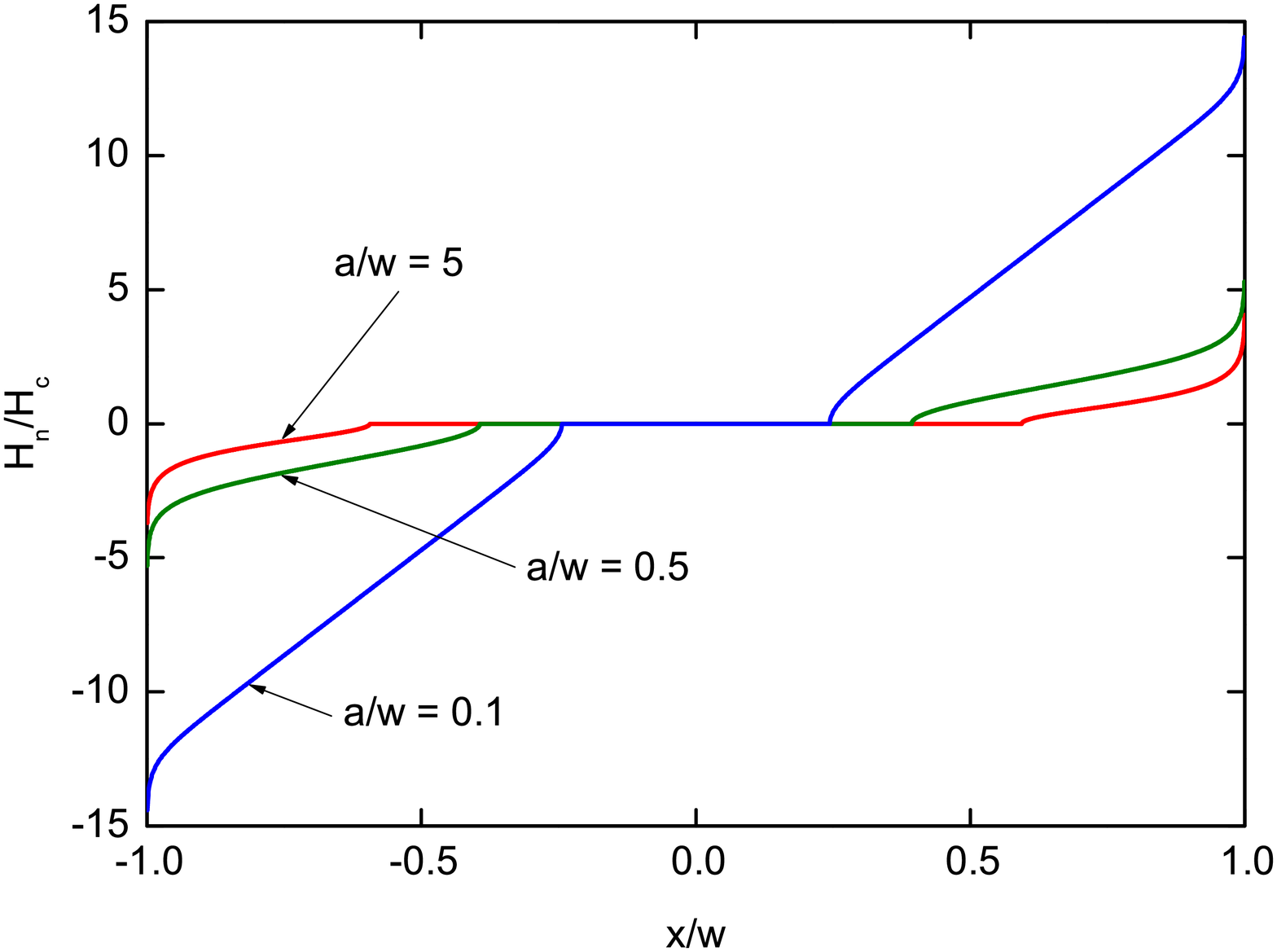} \hfill
    \includegraphics[width=8cm]{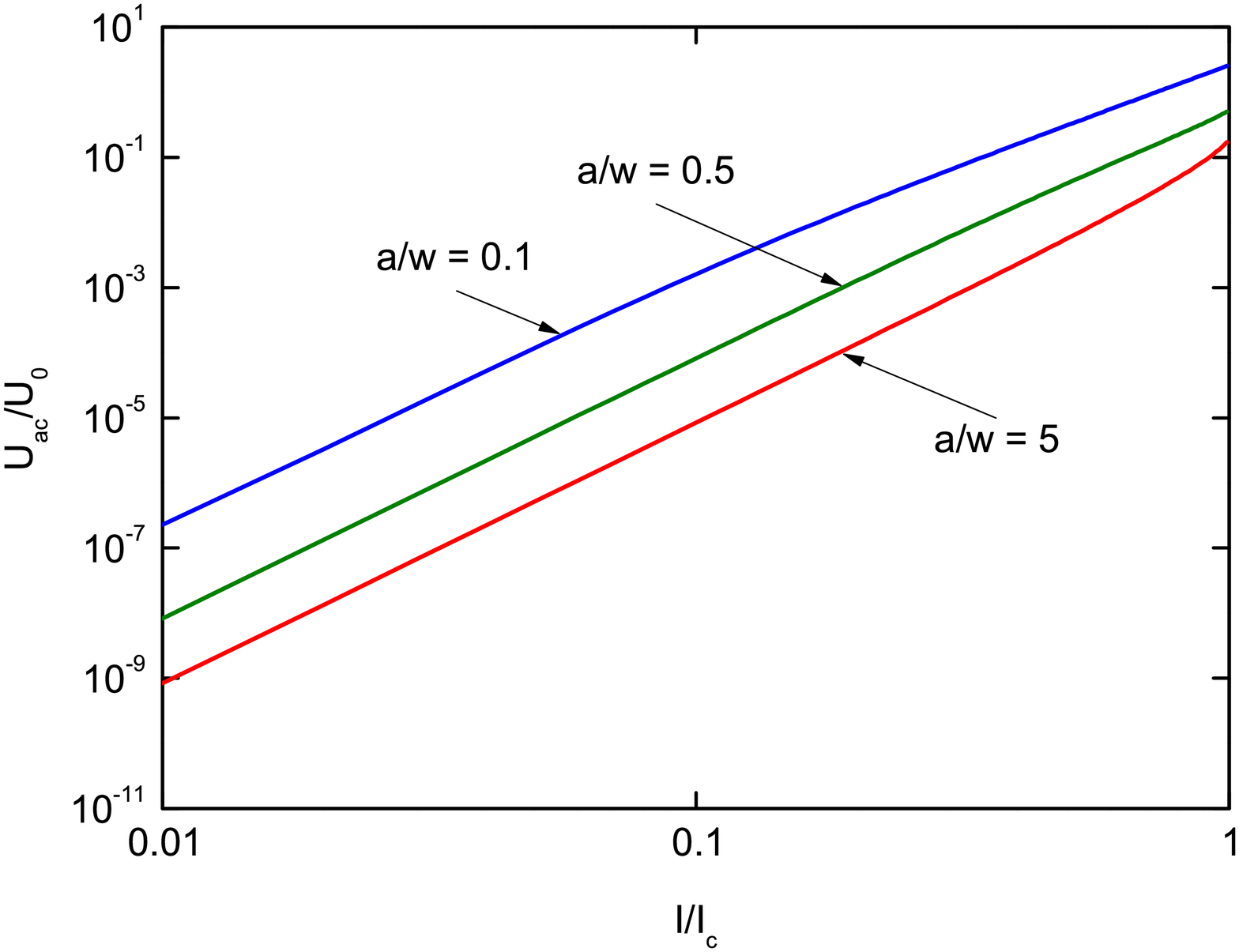}     \\
\parbox[t]{0.47\textwidth}{ \caption{(Colour online) Distribution of the normalized 
             component of the magnetic field $H_n/H_c$ over the width of the magnetically 
             shielded superconductor strip for three different values of the normalized 
             distance $a/w$ identified on the curves, when the normalized current 
             $I/I_c=0.8$, referring to the longitudinal shielding geometry of 
             Fig.~\ref{skizze}(a).}\label{cur-par} }\hfill
\parbox[t]{0.47\textwidth}{ \caption{(Colour online) Variation of the normalized 
             hysteretic ac loss $U_{ac}/U_0$ suffered by the magnetically shielded 
             superconductor strip with the normalized transport current $I/I_c$ 
             for three different values of the normalized  distance  $a/w$ identified 
             on the curves, referring to the longitudinal shielding geometry of 
             Fig.~\ref{skizze}(a).}\label{ac-par} }    
\end{figure*}

The variation of the normalized component of the magnetic field $H_n/H_c$ over the width 
of the strip, calculated from Eq.~(\ref{field-par}) for a range of values of the 
normalized vertical distance between the surfaces of the magnets and the centre of the 
strip  $a/w$, adopting the normalized current  $I/I_c=0.8$, is displayed in 
Fig.~\ref{cur-par}. Evidently, while at large $a/w$, the field profile virtually 
reproduces that of an isolated strip~\cite{Norris1970}, it straightens and augments in 
strength while showing a reduced flux-free zone, as $a/w$ abates, reminiscent of the 
field distribution in an isolated superconductor slab~\cite{Campbell}, for which the 
strip together with its magnetic mirror images effectively assembles into a stack of 
superconductor films~\cite{Mawatari1996}.

Substituting the normal component of the magnetic field, Eq.~(\ref{field-par}), into 
Eq.~(\ref{ac-loss}) and changing the variables $x$ and $x'$ to $u=\tanh{(\pi x/2a)}$   
and  $u'=\tanh{(\pi x'/2a)}$, respectively, yields the energy dissipated during a cycle 
of the ac transport current, per unit length of the strip, for the longitudinal shielding 
geometry,
\begin{align}
\label{Uac-par}
U_{ac} & = U_0  \left(\frac{2a}{\pi w }\right)^2 \nonumber\\
& \times\int_p^q \frac{du}{1-u^2} \int_p^{u}\frac{du'}{1-u'^2}\, 
\text{arctanh}\, \sqrt{ \frac{u'^2-p^2}{q^2-p^2} },
\end{align} 
where  $U_0=2\mu_0 I_c^2/\pi$. In the limit of the magnets situated close to the strip,  
$d\leq a\ll w$, Eq.~(\ref{Uac-par}) may be approximated with high accuracy by the 
expression 
\begin{equation}
\label{Uac-par-apr1} 
U_{ac} \simeq U_0 \left[ \frac{\pi w}{12a} \left( \frac{l}{w}\right)^3 
+ \frac{\ln{2}}{2}  \left( \frac{l}{w}\right)^2\right], 
\end{equation}
introducing the flux penetration depth $l=w-b$ , where $b$ is related to $p$ from 
Eq.~(\ref{p(I)}) as given above. A basically equivalent representation trying the current
$I$, albeit confined to the range $d/w\leq a/w\leq I/I_c\leq 1-a/w$, obtains from 
Eq.~(\ref{Uac-par-apr1}), since under these circumstances the approximations
\begin{gather}
\label{param-par}
p  \simeq 1 - \frac{1}{2} \exp{\left[-\frac{\pi w}{a}\left(1-\frac{I}{I_c}
\right)\right]},\, 
q \simeq 1-2 \exp{\left( -\frac{\pi w}{a} \right)  },\nonumber\\
b  \simeq a\left( \frac{2\ln{2}}{\pi}\right) + w\left(1-\frac{I}{I_c}\right)     
\end{gather} 
hold, so that the simple form
\begin{equation}
\label{Uac-par-apr2} 
U_{ac} \simeq U_0 \left( \frac{2a}{\pi w} \right)^2 f\left( \frac{\pi w}{2a} 
\frac{I}{I_c} \right),\, f(i)=\frac{1}{6}i^3-\frac{(\ln{2})^2}{2}i 
\end{equation}
ensues. The cubic term herein, which dominates in the high-current regime, describes the 
hysteretic ac loss as for an isolated superconductor slab\cite{Campbell}.

The dependence of the normalized hysteretic ac loss $U_{ac}/U_0$ on the normalized 
transport current $I/I_c$, calculated numerically from Eq.~(\ref{Uac-par}) for the above 
range of values of the normalized distance $a/w$, is portrayed in Fig.~\ref{ac-par}. 
This reveals that, while at large $a/w$, the ac loss is practically like for an 
unshielded strip~\cite{Norris1970}, it increases substantially, as $a/w$  abates, 
displaying a current variation like for an isolated superconductor slab\cite{Campbell}, 
the predictions of Eq.~(\ref{Uac-par-apr2}) differing indiscernibly within its
limitations.

\subsection{\label{subsec:quer} Transverse shielding geometry}

Following previous analysis, for the bulk flat magnets oriented perpendicular to the 
plane of the strip as shown in Fig.~\ref{skizze}(b), the normal component of the magnetic 
field in the flux-filled margins of the strip, $-w\leq x\leq b$ and $b\leq x\leq w$, 
reads~\cite{Genenko2000}
\begin{equation}
\label{field-quer}
H_n(x) =H_c\, \mbox{sgn}(x)\, \text{arctanh}\, \sqrt{ \frac{v^2(x)-r^2}{s^2-r^2} },   
\end{equation}
where $H_c=J_c/\pi$, apart from $v(x)=\tan{(\pi x/2a)}$, $r=\tan{(\pi b/2a)}$ and 
$s=\tan{(\pi w/2a)}$. Herein, since the half-width of the flux-free zone $b$ depends on 
the transport current $I$ itself,  
\begin{equation}
\label{r(I)}
r = \sqrt{\cos^2{\left( \frac{\pi w}{2a}\frac{I}{I_c} \right)} 
\sec^2{\left( \frac{\pi w}{2a} \right)} -1}.
\end{equation}

\begin{figure*}[tb]
    \includegraphics[width=8cm]{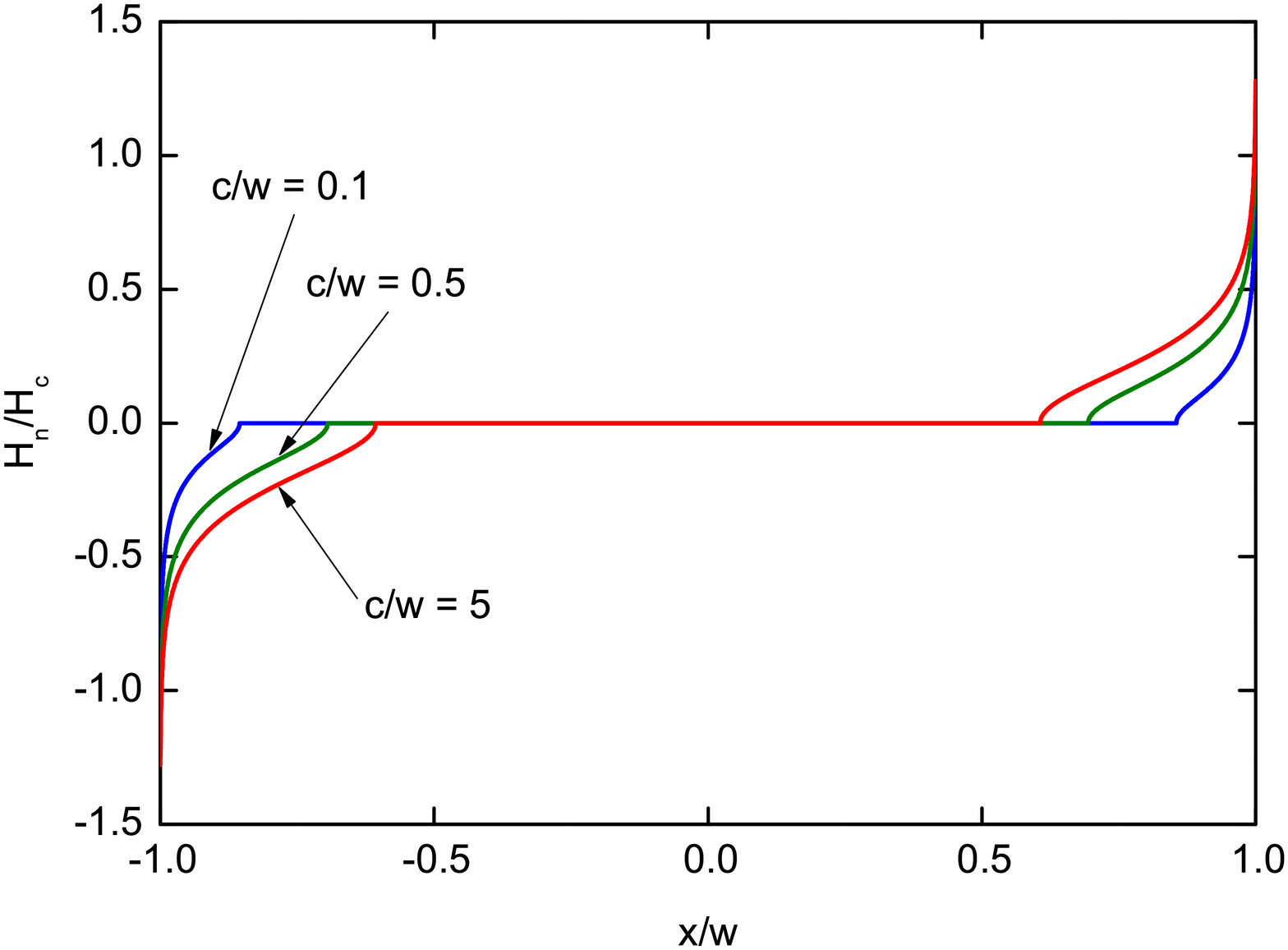} \hfill
    \includegraphics[width=8cm]{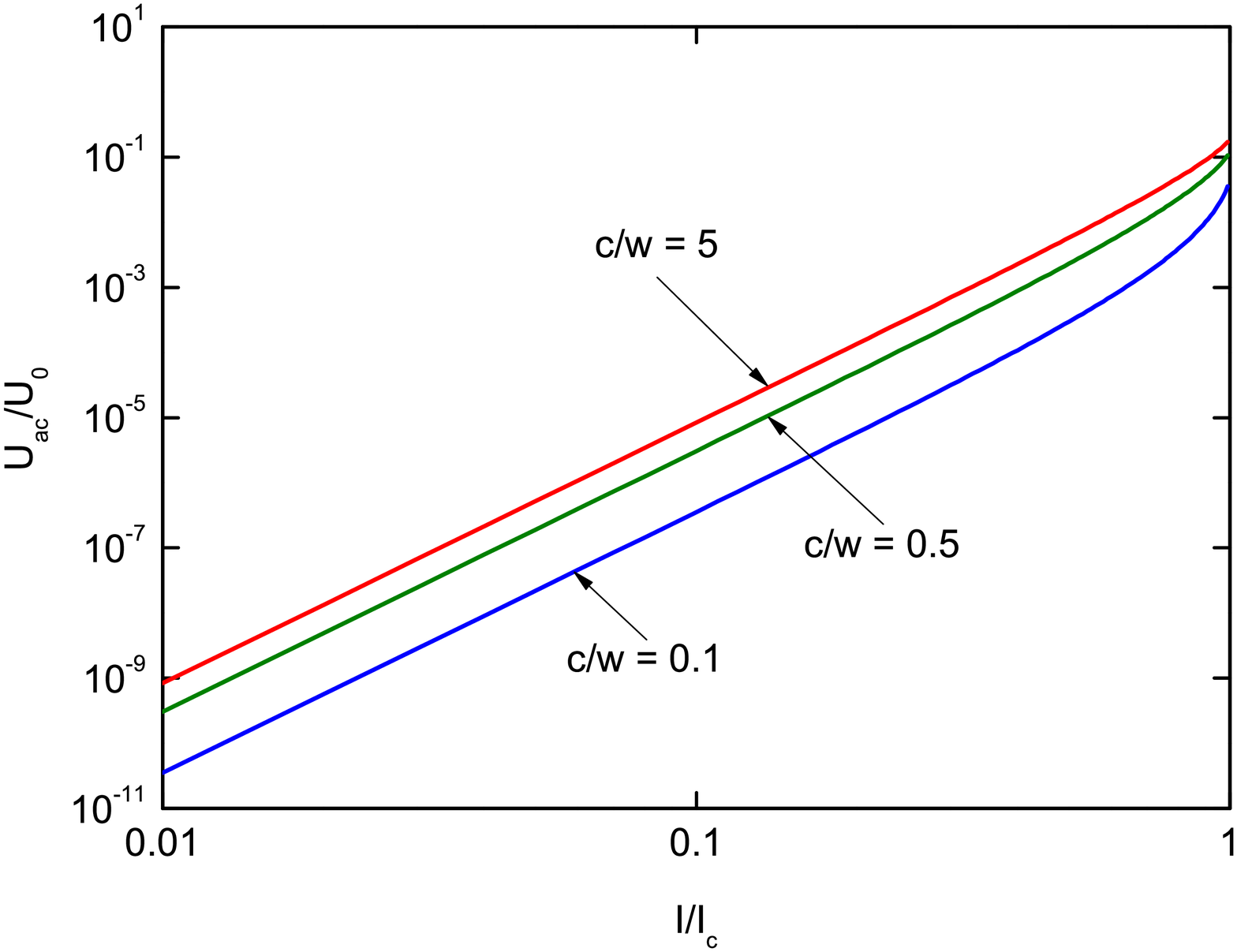}     \\
\parbox[t]{0.47\textwidth}{ \caption{(Colour online) Distribution of the normalized 
                          component of the magnetic field $H_n/H_c$ over the width of the 
                          magnetically shielded superconductor strip for three different 
                          values of the normalized distance $c/w$ identified on the curves, 
                          when the normalized current $I/I_c=0.8$, referring to the transverse 
                          shielding geometry of Fig.~\ref{skizze}(b).}\label{cur-perp} }
                          \hfill
\parbox[t]{0.47\textwidth}{ \caption{(Colour online) Variation of the normalized 
                          hysteretic ac loss $U_{ac}/U_0$ suffered by the magnetically 
                          shielded superconductor strip with the normalized transport 
                          current $I/I_c$ for three different values of the normalized  
                          distance $c/w$ identified on the curves, referring to the 
                          transverse shielding geometry of Fig.~\ref{skizze}(b).}\label{ac-perp} }    
\end{figure*}

The variation of the normalized component of the magnetic field $H_n/H_c$ over the width 
of the strip, calculated from Eq.~(\ref{field-quer}) for a range of values of the 
horizontal distance between the surfaces of the magnets and the edges of the strip, 
$c=a-w$, normalized by the half-width of the strip $w$, adopting the normalized 
current $I/I_c=0.8$, is displayed in Fig.~\ref{cur-perp}. Evidently, while at large 
$c/w$, the field profile virtually reproduces that of an isolated 
strip~\cite{Norris1970}, it steepens and weakens in strength while showing an enlarged 
flux-free zone, as $c/w$ abates, reminiscent of the field distribution in a strip located 
between two parallel superconducting shields~\cite{Genenko2002_2}.  

Substituting the normal component of the magnetic field, Eq.~(\ref{field-quer}), into 
Eq.~(\ref{ac-loss}) and changing the variables $x$ and $x'$ to $v=\tan{(\pi x/2a)}$  
and $v'=\tan{(\pi x'/2a)}$, respectively, yields the energy dissipated during a cycle 
of the ac transport current, per unit length of the strip, for the transverse shielding 
geometry,
\begin{align}
\label{Uac-quer}
U_{ac} & = U_0  \left(\frac{2a}{\pi w }\right)^2 \nonumber\\
& \times \int_r^s \frac{dv}{1+v^2} \int_r^{v}\frac{dv'}{1+v'^2}\, 
\text{arctanh}\, \sqrt{ \frac{v'^2-r^2}{s^2-r^2} },
\end{align}
where  $U_0=2\mu_0 I_c^2/\pi$. In the limit of the magnets situated close to the strip,  
$d\leq c\ll w$, Eq.~(\ref{Uac-quer}) may be approximated with high accuracy by the 
expression 
\begin{align}
\label{U(l)1}
U_{ac} & \simeq U_0  \left(\frac{c}{w}\right)^2\\ 
& \times \left[ \sqrt{ \frac{l}{2c} \left(\frac{l}{2c} +1 \right) }\,
\mbox{arcsec}\left(\frac{l}{c} +1 \right)
- \ln{ \left(\frac{l}{c} +1 \right) } \right],\nonumber
\end{align}
introducing the flux penetration depth $l=w-b$, where $b$ is related to $r$ from 
Eq.~(\ref{r(I)}) as given above. A basically equivalent representation trying the 
current $I$, albeit confined to the range $0\leq I/I_c\leq 1-c/w$, obtains from 
Eq.~(\ref{U(l)1}), since under these circumstances the approximations
\begin{gather}
\label{param-quer}
r \simeq s\cos{ \left( \frac{\pi w}{2a} \frac{I}{I_c} \right) },\,
s  \simeq \frac{2a}{\pi c} \left[1 -\frac{\pi^2}{12} \left(\frac{c}{w}\right)^2 
\right]  ,\nonumber\\
b \simeq a-c\sec{ \left(\frac{\pi w}{2a}\frac{I}{I_c}\right)}
\end{gather} 
hold, so that the simple form
\begin{equation}
\label{Uac-quer-apr} 
U_{ac} \simeq U_0 \left( \frac{c}{w} \right)^2 
f\left( \frac{\pi w}{2a}\frac{I}{I_c} \right),\,
f(i)=\frac{1}{2}i \tan{i} +\ln{\cos{i}}
\end{equation} 
ensues. The quadratic prefactor herein describes the hysteretic ac loss as for a 
superconductor strip between two parallel superconducting shields~\cite{Genenko2004}.

The dependence of the normalized hysteretic ac loss $U_{ac}/U_0$ on the normalized 
transport current $I/I_c$, calculated numerically from Eq.~(\ref{Uac-quer}) for the 
above range of values of the normalized distance $c/w$, is portrayed in 
Fig.~\ref{ac-perp}. This reveals that, while at large $c/w$, the ac loss is practically 
like for an unshielded strip~\cite{Norris1970}, it decreases substantially, as $c/w$ 
abates, displaying a current variation like for a strip between two parallel 
superconducting shields~\cite{Genenko2004}, the predictions of Eq.~(\ref{Uac-quer-apr}) 
differing indiscernibly within its limitations.

\section{\label{sec:disc} Summary and conclusion} 

Based on a quasistatic approach, we have presented exact numerical calculations and 
approximate analytic forms delineating the penetration of magnetic flux and hysteretic 
ac losses in a thin, current-carrying type-II superconductor strip located between two 
flat magnetic shields. For the shields oriented parallel or, 
respectively, perpendicular to the plane of the strip, our results predict a possible 
strong increase or, respectively, decrease of the hysteretic loss, when the shields 
approach the strip. The simple analytic forms derived on the assumption of infinite 
permeability can serve as guides for estimating ac losses in practically relevant 
configurations involving magnetic shields of finite permeability too.

\bibliographystyle{plain}
\bibliography{apssamp}

\end{document}